\documentclass[10pt,conference]{IEEEtran}   
\usepackage{graphics} 
\usepackage{epsfig}
\usepackage{amsmath} 
\usepackage{amssymb}  
\usepackage{url}
\usepackage[ruled, vlined, linesnumbered]{algorithm2e}
\usepackage{verbatim} 
\usepackage{soul, color}
\usepackage{lmodern}
\usepackage{fancyhdr}
\usepackage[utf8]{inputenc}
\usepackage{fourier} 
\usepackage{array}
\usepackage{makecell}
\usepackage{fancyhdr}
\usepackage{listings}
\usepackage{graphicx}
\usepackage{multirow}
\usepackage{pdfpages}

\title{\textit{CPPJoules}: An Energy Measurement Tool for C++}

\begin{document}
\graphicspath{ {./images/} }

\author{
Shivadharshan S, Akilesh P, Rajrupa Chattaraj, Sridhar Chimalakonda\\
\textit{Research in Intelligent Software and Human Analytics (RISHA) Lab} \\
Department of Computer Science and Engineering \\
Indian Institute of Technology Tirupati, India \\
\{cs22b057, cs22b040, cs22s504, ch\}@iittp.ac.in
}

\maketitle

\lstset{
  numbers=left,       
  stepnumber=1,       
  numbersep=5pt,     
  frame=single,       
  basicstyle=\ttfamily\small,  
  commentstyle=\color{green},   
  numberstyle=\normalsize,
  xleftmargin=10pt,
  framexleftmargin=10pt
}

\begin{abstract}
 
 With the increasing complexity of modern software and the demand for high performance, energy consumption has become a critical factor for developers and researchers. While much of the research community is focused on evaluating the energy consumption of machine learning and artificial intelligence systems—often implemented in Python—there is a gap when it comes to tools and frameworks for measuring energy usage in other programming languages. C++, in particular, remains a foundational language for a wide range of software applications, from game development to parallel programming frameworks, yet lacks dedicated energy measurement solutions. To address this, we have developed \textit{CPPJoules}, a tool built on top of Intel-RAPL to measure the energy consumption of C++ code snippets. We have evaluated the tool by measuring the energy consumption of the standard computational tasks from the Rosetta Code repository. The demonstration of the tool is available at \url{https://www.youtube.com/watch?v=GZXYF3AKzPk} and related artifacts at \url{https://rishalab.github.io/CPPJoules/}.  
 \\
\end{abstract}

\section{INTRODUCTION}

Energy consumption has become a critical concern in software development, particularly for performance-critical and resource-intensive applications. As computational workloads grow in complexity, the demand for tools that provide energy consumption metrics has intensified. These tools enable developers to design energy-efficient software by understanding the energy cost associated with various code segments. Although hardware-based interfaces like Intel-RAPL offer a way to measure energy consumption, they often require considerable effort. To address these challenges, tools such as pyJoules\footnote{\url{https://pyjoules.readthedocs.io/}}, CodeCarbon\footnote{\url{https://codecarbon.io}}, jRAPL\cite{liu2015data}, and RJoules\cite{chattaraj2023rjoules} facilitate energy consumption measurement for different programming languages. However, despite the availability of energy measuring tools for languages such as Python, Java, and R, there remains a gap in  energy measurement tools for C++, a language widely used in performance-critical applications.


C++ is employed in fields where computational workloads are substantial, such as high-performance computing, real-time systems, system programming, embedded systems, and game development \cite{andrist2020c++}. Approximately 3.2 million public repositories on GitHub use C++ as their primary programming language\footnote{\url{https://api.github.com/search/repositories?q=language:C++}}, reflecting its widespread adoption in various domains. Many machine learning and deep learning frameworks such as XGBoost\footnote{\url{https://xgboost.readthedocs.io/en/latest/}}, TensorFlow\footnote{\url{https://www.tensorflow.org}}, and PyTorch\footnote{\url{https://pytorch.org}} are built on C++ due to its efficiency and close-to-hardware control features. Moreover, CUDA\footnote{\url{https://developer.nvidia.com/cuda-toolkit}}, the leading parallel computing framework for GPU programming, is built on C++. These domains inherently demand significant energy resources due to their hardware-intensive operations.

Additionally, C++ is used for developing database engines and systems such as MySQL and MongoDB, where performance and control over resources are critical. Furthermore, many popular open-source development frameworks, such as Apache Hadoop, Node.js, and OpenCV, rely on C++ for core components due to its performance benefits. Given C++'s widespread use in performance-critical domains, understanding and managing energy consumption is crucial for developers and researchers aiming to optimize energy efficiency. To address this need, we propose \textbf{\textit{CPPJoules}}, a dedicated tool designed specifically to measure the energy footprint of a host machine during the execution of a C++ code snippet.

 \textit{CPPJoules} leverages the Intel Running Average Power Limit (RAPL) interface, a well-established mechanism for reporting energy consumption of different system-on-chip (SoC) power domains. RAPL has been extensively validated in independent research\cite{babakol2024tensor,nahrstedt2024empirical}, for energy measurement in modern CPUs. \textit{CPPJoules} utilizes the \texttt{powercap} interface on Linux to access energy consumption data directly from various domains of Intel-RAPL. For cross-platform support, \textit{CPPJoules} utilizes \texttt{intel-power-gadget}\footnote{\url{https://github.com/gmierz/intel-power-gadget#}} on Windows operating systems, which fetches energy data from Intel-RAPL via the Model Specific Registers (MSR). Additionally, for GPU energy measurement, \textit{CPPJoules} leverages \texttt{NVML}\footnote{\url{https://developer.nvidia.com/management-library-nvml}} library, a C-based API for monitoring and managing various states of the NVIDIA GPU devices. This provides direct access to the queries and commands exposed via \texttt{nvidia-smi}\footnote{\url{https://developer.nvidia.com/system-management-interface}}, similar to the method used in \cite{rajput2024greenlight}.


We evaluated \textit{CPPJoules} by benchmarking it against the energy trends of pyJoules on a standard set of tasks, including both CPU and GPU-intensive workloads, as detailed in Section \ref{evaluation}. Our evaluation validated the consistency of energy measurement trends and run-time performance, confirming the accuracy and reliability of the tool in monitoring energy consumption during C++ code execution. 

\begin{figure*}[htbp]
    \centering
    \makebox[\textwidth][c]{\includegraphics[width=0.87\paperwidth]{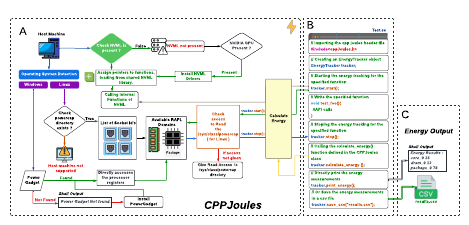}}
    \caption{Execution Flow Diagram of \textit{CPPJoules}: A illustrates the internal workflow, B presents an example code snippet for integration, and C shows the output derived from \textit{CPPJoules}.}
    \label{fig:cppjoules}
    \vspace{-2mm}
\end{figure*}

\begin{table*}[t]
    \begin{center}
    \begin{tabular}{|c c c c c c c c c c c c c|}
        \hline
        & \multicolumn{3}{c}{\textbf{PKG Energy}} & \multicolumn{3}{c}{\textbf{DRAM Energy}} & \multicolumn{3}{c}{\textbf{GPU Energy}} & \multicolumn{3}{c|}{\textbf{Run-Time (s)}} \\
        \hline
        \textbf{Tasks} & \textbf{Python} & \textbf{C++} & \textbf{p-value} & \textbf{Python} & \textbf{C++} & \textbf{p-value} & \textbf{Python} & \textbf{C++} & \textbf{p-value} & \textbf{Python} & \textbf{C++} & \textbf{p-value} \\
        \hline
        array\_concat & 869.724 & 137.396 & 0.002 & 71.955 & 12.020 & 0.002 & 0.092 & 0.014 & 0.002 & 6.89 & 1.11 & 0.002 \\
        bubble\_sort & 1279.181 & 164.272 & 0.004 & 94.835 & 12.222 & 0.004 & 0.137 & 0.017 & 0.004 & 10.23 & 1.32 & 0.004 \\
        insertion\_sort & 386.458 & 26.583 & 0.002 & 29.376 & 1.995 & 0.002 & 0.041 & 0.002 & 0.002 & 3.09 & 0.22 & 0.002 \\
        merge\_sort & 7903.322 & 839.029 & 0.002 & 637.656 & 64.961 & 0.002 & 0.833 & 0.089 & 0.002 & 62.66 & 6.7 & 0.002 \\
        n\_body & 41596.306 & 14959.442 & 0.008 & 3381.634 & 1148.572 & 0.008 & 4.365 & 1.571 & 0.008 & 326.24 & 117.22 & 0.008 \\
        matrix\_mul & 11429.731 & 2773.890 & 0.008 & 872.205 & 215.519 & 0.008 & 1.196 & 0.295 & 0.008 & 89.44 & 22.07 & 0.008 \\
        \hline
    \end{tabular}
    \end{center}
    \caption{Mean values of Energy Consumption (JOULES), Run-Time (SECONDS), and Wilcoxon p-value Comparison between Python and C++.}
    \label{table:eval}
    \vspace{-6mm}
\end{table*}

\section{DESIGN \& DEVELOPMENT}

\subsection{Back-end Framework}
\label{sec:back-end}

\textit{CPPJoules} leverages Intel's RAPL and NVIDIA's \texttt{NVML} libraries to measure energy consumption for both CPU and GPU components. For CPU energy measurement, it accesses detailed energy data across various domains using the \texttt{powercap} interface on Linux and \texttt{intel-power-gadget} on Windows.

\begin{itemize}
    \item \textbf{\texttt{powercap}:}  A Linux kernel subsystem that interfaces with Intel-RAPL to provide energy consumption data for various processor domains such as package, core, uncore and DRAM. The package domain measures total energy usage for the CPU socket, including cores and integrated components. The core domain tracks energy used by CPU cores, while the uncore domain covers components outside the cores, such as memory controllers and GPUs. The DRAM domain measures energy used by the system's RAM.
    


     \item \textbf{\texttt{intel-power-gadget}:} A Windows tool for energy monitoring of Intel processors, tracking CPU cores, memory, and integrated graphics. It reports various domains: processor energy for the entire CPU, DRAM for system memory (if enabled in BIOS), IA for CPU cores, and GT for integrated graphics.
     

\end{itemize}


For GPU energy measurement, \textit{CPPJoules} integrates \texttt{NVML} (NVIDIA Management Library), using the API function \texttt{nvmlDeviceGetTotalEnergyConsumption()} to capture the total energy during specific tasks.

\subsection{Execution Flow of \textit{CPPJoules}}
This section describes the internal workflow of \textit{CPPJoules} on the host system, as depicted in Fig. \ref{fig:cppjoules}. The process begins with a verification step to ensure compatibility with the host system by locating the \texttt{powercap} directory. This directory is typically present on Linux machines equipped with modern Intel processors, and its absence indicates that \textit{CPPJoules} will not function on the system.

On Linux, the necessary files are installed in the \texttt{/usr} directory, making them accessible for the system's compilers. Once compatibility is confirmed, \textit{CPPJoules} retrieves a list of available socket identifiers from Intel's RAPL directory, as illustrated in Fig. \ref{fig:cppjoules}.A, and fetches the corresponding power domains, as described in Section \ref{sec:back-end}. The tool iterates over the available socket IDs and power domains, requiring read access to the \texttt{/sys/class/powercap} directories to gather energy measurements. 

For systems running Windows, \textit{CPPJoules} checks for the presence of \texttt{intel-power-gadget}; if found, it retrieves the RAPL domains directly. Otherwise, users are prompted to install the required software. If the \texttt{NVML} library is not available, the tool will continue to operate, but it will only report energy measurements related to the CPU processor domains. This ensures that energy tracking can still be performed for the CPU, even if GPU monitoring is not supported on the system.
Energy monitoring begins with the \texttt{start()} method to take an initial reading and concludes with \texttt{stop()}, marking the end of tracking, as depicted in Fig. \ref{fig:cppjoules}.B. The tool then calculates the energy consumption by finding the difference between these readings, a commonly used approach in prior research \cite{pereira2017energy,chattaraj2023rjoules}.

To measure the energy consumption of a C++ code snippet, the user first needs to include the header file of \textit{CPPJoules}, as depicted in Fig. \ref{fig:cppjoules}.B, and create an instance of the \texttt{EnergyTracker} class (referred to as tracker), which serves as the core component for the energy measurement process. Upon initialization, the \texttt{EnergyTracker} object scans the system for the required libraries and directories needed for energy tracking, as shown in Fig. \ref{fig:cppjoules}.A.

To perform the measurement, the user invokes the \texttt{tracker.start()} method, followed by placing the C++ code snippet or function call to be measured, and then calls \texttt{tracker.end()} to conclude the tracking. For finer granularity, the tool supports repeated calls to \texttt{start()} and \texttt{stop()} over short intervals, reducing the measurement error caused by the RAPL counter's overflow during longer executions \cite{khan2018rapl}. The total energy consumption and run-time can be calculated using the \texttt{calculate\_energy()} method. Users can choose to display the energy measurements with \texttt{print\_energy()} or save the results to a \texttt{CSV} file using the \texttt{save\_csv()} method, as illustrated in Fig. \ref{fig:cppjoules}.B. Below is an example demonstrating the use of \textit{CPPJoules} within a C++ code snippet:
\begin{lstlisting}[language=c++, frame=single]
#include <cppJoules.h> 
EnergyTracker tracker; 
tracker.start(); 
// Place your code or function calls here
tracker.stop(); 
tracker.calculate_energy(); 
tracker.save_csv("file_name.csv"); 
\end{lstlisting}

\vspace{-3mm}
\section{EVALUATION}
\label{evaluation}

For evaluation, we do not have an established ground truth for energy consumption measurements. Since \textit{CPPJoules} is the first tool specifically designed for measuring energy consumption in C++ programs, we adopted a methodology similar to \textit{RJoules}\cite{chattaraj2023rjoules}, where a set of standard computational tasks is used to validate energy measurements based on observed trends. To assess the accuracy of \textit{CPPJoules}, we conducted an empirical experiment by benchmarking it against the energy trends observed from a widely used energy measurement tool, pyJoules\footnote{\url{https://pyjoules.readthedocs.io/}}, which relies on Intel's RAPL interface. The experiment compared energy consumption measurements for Python (using pyJoules) and C++ (using \textit{CPPJoules}) across the same tasks.

We utilized the RosettaCode repository\footnote{\url{https://github.com/acmeism/RosettaCodeData}}, a publicly available programming chrestomathy site that offers a comprehensive set of tasks, including 851 completed tasks and 230 draft tasks, spanning 658 different programming languages. However, many of these tasks are small, making them unsuitable for accurate energy measurement due to Intel RAPL's low sampling rate of around 1 ms (1,000 samples per second)\cite{khan2018rapl}. To ensure reliable results, we selected a subset of CPU- and GPU-intensive tasks, such as array concatenation, bubble sort, insertion sort, merge sort, n-body simulation, and matrix multiplication, as identified by Georgiou et al. \cite{georgiou2017analyzing}.


The experiments were conducted on an Ubuntu system equipped with an Intel(R) Xeon(R) Gold 6226R CPU, running at a frequency of 2.90 GHz, with 16 cores and 128 GB of RAM. To maintain consistency and minimize external interference, identical input data was used across all tasks. Additionally, non-essential background processes were manually terminated before the start of each experiment to reduce noise and variability in the measurements.

To mitigate the effects of system noise and transient interference on energy measurements, each task was executed ten times, and the average energy consumption was recorded. This methodology is consistent with the approach outlined by Georgiou et al. \cite{georgiou2022green, shanbhag2023exploratory}. Furthermore, to prevent the effect of power tail states\cite{bornholt2012model}, we introduced a 30-second idle period between iterations. The energy readings and run-time values from \textit{CPPJoules} and pyJoules were saved in CSV files after each iteration.

\begin{figure}[ht]
    \centering
    \includegraphics[scale=0.38]{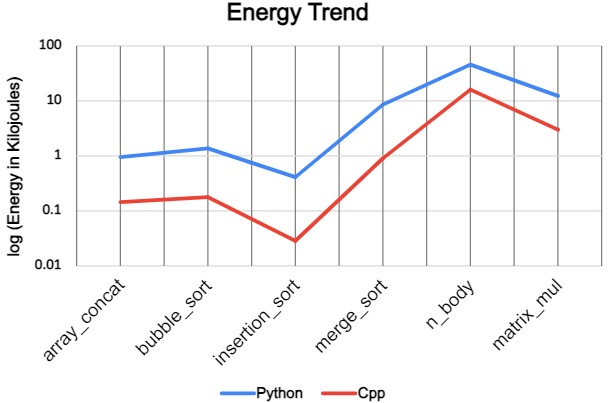}
    \caption{Mean Energy Consumption for Python and C++ Across Examined Tasks}
    \label{fig:energy_trend}
    \vspace{-4mm}
\end{figure}

Table \ref{table:eval} presents the mean values of package (PKG), DRAM, and GPU energy consumption (in Joules and rounded to 3 decimals for readability), along with the average run-time for each task. The results show that, for both Python and C++, energy consumption and run-time follow similar trends, with tasks ranked in ascending order as insertion sort, array concatenation, bubble sort, merge sort, matrix multiplication, and n-body simulation. This consistency supports the validity of \textit{CPPJoules} measurements compared to pyJoules.
To further illustrate these findings, Fig \ref{fig:energy_trend} shows energy consumption for both Python and C++, reinforcing the patterns observed in Table \ref{table:eval} and confirming the reliability of \textit{CPPJoules} in energy measurement.

To determine whether there is a statistically significant difference between the raw outputs—specifically, the actual energy consumption readings, rather than their mean values—we conducted the Wilcoxon Signed Rank Test \cite{wilcoxon1992individual} on all four output variables: PKG energy, DRAM energy, GPU energy, and run-time. The resulting p-values, shown in Table \ref{table:eval}, were all below 0.05, indicating \textit{no statistically significant difference} between the outcomes measured by \textit{CPPJoules} and pyJoules. This suggests that both tools produce comparable results in terms of energy and performance measurement. For further analysis and transparency, the evaluation scripts and detailed energy consumption data for each trial are publicly available at: \url{https://github.com/rishalab/CPPJoules/tree/main/evaluation}.


\section{LIMITATION \& FUTURE WORK}
\textit{CPPJoules} is built on top of Intel RAPL, making it currently limited to systems with Intel processors, specifically those from the Celeron, Pentium, Core i3, i7, i9, and Xeon series with the Sandy Bridge or newer architecture. The current version of the tool integrates additional packages such as \texttt{powercap} and \texttt{intel-power-gadget}. However, since the \texttt{intel-power-gadget} library has been discontinued, the next version of the tool will incorporate the Intel Performance Counter Monitor\footnote{\url{https://github.com/intel/pcm}} for enhanced functionality.

The current version of the tool supports both Linux and Windows operating systems, with plans to extend support to macOS in future updates. Additionally, since \textit{CPPJoules} uses \texttt{nvidia-smi} for GPU energy measurements, it currently supports only NVIDIA GPUs. Future versions of the tool will aim to include support for AMD GPUs, broadening its applicability in diverse hardware environments.






\section{CONCLUSION}

In conclusion, we present \textit{CPPJoules}, a tool designed to measure energy consumption in C++ code snippets. Through our experiments using standard computational tasks, we have demonstrated its accuracy and utility in profiling the energy consumption.



\textit{CPPJoules} offers significant potential for advancing research into the energy consumption of C++ libraries, such as the Standard Template Library (STL), and many open source frameworks which internally uses C++ at their core components. \textit{CPPJoules} provides a platform for conducting in-depth studies on the energy efficiency of such frameworks, helping Software Engineering (SE) researchers and developers better understand energy consumption patterns at the library and framework level. 

Our findings highlight the importance of developing energy measurement tools that cater specifically to the nuances of compiled languages like C++, which often exhibit different energy profiles compared to interpreted languages. As energy consumption becomes an increasingly critical factor in software development, tools like \textit{CPPJoules} will play a crucial role in promoting energy-efficient coding practices and further optimizing the performance of libraries and frameworks developed using C++.



\addtolength{\textheight}{-12cm}   






\bibliographystyle{ieeetr}
\bibliography{references}

\begin{thebibliography}{10}

\bibitem{liu2015data}
K.~Liu, G.~Pinto, and Y.~D. Liu, ``Data-oriented characterization of application-level energy optimization,'' in {\em Fundamental Approaches to Software Engineering: 18th International Conference, FASE 2015, Held as Part of the European Joint Conferences on Theory and Practice of Software, ETAPS 2015, London, UK, April 11-18, 2015, Proceedings 18}, pp.~316--331, Springer, 2015.

\bibitem{chattaraj2023rjoules}
R.~Chattaraj and S.~Chimalakonda, ``Rjoules: An energy measurement tool for r,'' in {\em 2023 38th IEEE/ACM International Conference on Automated Software Engineering (ASE)}, pp.~2026--2029, IEEE, 2023.

\bibitem{andrist2020c++}
B.~Andrist, V.~Sehr, and B.~Garney, {\em C++ high performance: master the art of optimizing the functioning of your C++ code}.
\newblock Packt Publishing Ltd, 2020.

\bibitem{babakol2024tensor}
T.~Babakol and Y.~D. Liu, ``Tensor-aware energy accounting,'' in {\em Proceedings of the IEEE/ACM 46th International Conference on Software Engineering}, pp.~1--12, 2024.

\bibitem{nahrstedt2024empirical}
F.~Nahrstedt, M.~Karmouche, K.~Bargie{\l}, P.~Banijamali, A.~Nalini Pradeep~Kumar, and I.~Malavolta, ``An empirical study on the energy usage and performance of pandas and polars data analysis python libraries,'' in {\em Proceedings of the 28th International Conference on Evaluation and Assessment in Software Engineering}, pp.~58--68, 2024.

\bibitem{rajput2024greenlight}
S.~Rajput, M.~Kechagia, F.~Sarro, and T.~Sharma, ``Greenlight: Highlighting tensorflow apis energy footprint,'' in {\em 2024 IEEE/ACM 21st International Conference on Mining Software Repositories (MSR)}, pp.~304--308, IEEE, 2024.

\bibitem{pereira2017energy}
R.~Pereira, M.~Couto, F.~Ribeiro, R.~Rua, J.~Cunha, J.~P. Fernandes, and J.~Saraiva, ``Energy efficiency across programming languages: how do energy, time, and memory relate?,'' in {\em Proceedings of the 10th ACM SIGPLAN international conference on software language engineering}, pp.~256--267, 2017.

\bibitem{khan2018rapl}
K.~N. Khan, M.~Hirki, T.~Niemi, J.~K. Nurminen, and Z.~Ou, ``Rapl in action: Experiences in using rapl for power measurements,'' {\em ACM Transactions on Modeling and Performance Evaluation of Computing Systems (TOMPECS)}, vol.~3, no.~2, pp.~1--26, 2018.

\bibitem{georgiou2017analyzing}
S.~Georgiou, M.~Kechagia, and D.~Spinellis, ``Analyzing programming languages' energy consumption: An empirical study,'' in {\em Proceedings of the 21st Pan-Hellenic Conference on Informatics}, pp.~1--6, 2017.

\bibitem{georgiou2022green}
S.~Georgiou, M.~Kechagia, T.~Sharma, F.~Sarro, and Y.~Zou, ``Green ai: Do deep learning frameworks have different costs?,'' in {\em Proceedings of the 44th International Conference on Software Engineering}, pp.~1082--1094, 2022.

\bibitem{shanbhag2023exploratory}
S.~Shanbhag and S.~Chimalakonda, ``An exploratory study on energy consumption of dataframe processing libraries,'' in {\em 2023 IEEE/ACM 20th International Conference on Mining Software Repositories (MSR)}, pp.~284--295, IEEE, 2023.

\bibitem{bornholt2012model}
J.~Bornholt, T.~Mytkowicz, and K.~S. McKinley, ``The model is not enough: Understanding energy consumption in mobile devices,'' in {\em 2012 IEEE Hot Chips 24 Symposium (HCS)}, pp.~1--3, IEEE, 2012.

\bibitem{wilcoxon1992individual}
F.~Wilcoxon, ``Individual comparisons by ranking methods,'' in {\em Breakthroughs in statistics: Methodology and distribution}, pp.~196--202, Springer, 1992.

\end{thebibliography}
\end{document}